\documentclass[aps,showpacs,showkeys,preprint,nofootinbib]{revtex4}
\begin{document}
\newcommand{\be}{\begin{equation}}
\newcommand{\ee}{\end{equation}}
\bibliographystyle{unsrt}
\title{Stationary flows in quantum dissipative closed circuits as a
challenge to thermodynamics}
\author{V. \v{C}\'{a}pek and J. Bok}
\affiliation {Institute of Physics of Charles University, Faculty
of Mathematics and Physics, Ke Karlovu 5, 121 16 Prague 2, Czech
Republic\\(Tel. (**-420-2)2191-1330, Fax (**-420-2)2492-2797,
E-mail capek@karlov.mff.cuni.cz)}
\date{
April 24, 2002} 

\begin{abstract}

An experimentally inspired model is constructed and rigorously
solved from the Hamiltonian level where a dc circular spontaneous
flow exists in absence of a magnetic field, irrespective of
presence of dissipation causing otherwise proper relaxation. The
flow causes a spontaneous unidirectional transfer of heat from one
bath to another one, even against temperature step. This is what
is explicitly forbidden by the Clausius form of the Second law of
thermodynamics. The unidirectionality of the flow is caused by
that of spontaneous processes known to bear this property since
their introduction by Einstein. The model slightly improves the
previous one (\v{C}\'{a}pek \& Sheehan 2002), describes a
realistic plasma system for which experimental results violating
the second law were announced, and the result obtained fully
supports the experimental conclusions (Sheehan 1995). Analytical
proof of the violation is supported by numerical results. All
mathematical details are exposed, two fully independent types of
mathematical arguments behind starting equations are invoked, and
no approximations that could be made responsible for the striking
conclusions are used. It shows how the physics beyond the Second
law is still little understood.
\end{abstract} \vspace{0.25cm}

\maketitle
\newpage

\section{Introduction}

Challenges to the second law of thermodynamics \cite{Thom,Clau}
are almost as old as the law itself, dating back at least to the
1870's with Maxwell's celebrated demon
\cite{Maxw,Losc,Szil,LefRex}. Most of them have been resolved
under close scrutiny \cite{Szil,FeLeSa,Muse} but some persisted.
Anyway, strong belief in old authorities and natural human
tendency to organize things and facts into closed logical units
and complexes (scientific disciplines etc.) caused that almost
nobody doubts about validity of standard thermodynamics, in
particular the Second law, in at least the macroworld
\cite{LieYng1,LieYng2}. Absolutistic statements like `...No
exception to the second law of thermodynamics has ever been found
- not even a tiny one...' \cite{LieYng2} often appear and the
second law is almost universally believed to be unquestionable. In
our opinion, the situation is (in view of the fast developing
situation perhaps at least still) not so clear. The reason is that
experiments questioning the second law have been reported since
1995, have since been subject to a public discussion but remain so
far unquestioned \cite{Shee1,Niku}. Some of these experiments have
since been even reproduced (compare \cite{NikZhi} with
\cite{Niku2}).

In theory, arguments independent of these experiments appeared
since 1997 \cite{Cape1} saying that in quantum systems with strong
or at least intermediate coupling to its surroundings (identical
or connected with usual thermodynamic baths) with mutual strong
correlations (entanglement), the standard statistical
thermodynamics could be violated. This is in particular, but not
only, the case of the second law of thermodynamics. One must keep
in mind that from first principles (microscopic Hamiltonian
dynamics), derivations of the second law are declared to exist
just in classical (in the sense of non-quantum) physics
\cite{Mart} or, as in standard textbooks, involving assumption of
a weak system-bath coupling only. \footnote{Gibbs canonical form
of the system density matrix compatible with the standard
thermodynamics is correct to just the zeroth order in the
system-bath coupling. This is an important fact to be realized
already here as the effect reported below is of higher order in
this coupling.} It should be stressed already here that the
classical physics is, according to the Bohr correspondence
principle, an infinite temperature limit of the (more general)
quantum physics. This, inter alia, implies that its application to
finite temperatures as in standard thermodynamics is at least open
to discussion. The above lack of general derivation of the second
law beyond, in particular, the classical regime could also
correspond to the fact that so far reported and seemingly
classical paradoxes connected with the second law (see, e.g.,
\cite{Shepar}) usually involve sufficiently intense processes that
are inherently of the quantum character.

In 1999, it was realized that a long lasting call in chemistry for
inclusion of self-organizational tendencies into theory of
particle-transfer chemical reactions is, from the microscopic
point of view, nothing but a call for inclusion of such mechanisms
that can turn any (from the thermodynamical point of view) passive
bath into an active one, opening thus door to violations of
standard thermodynamic principles \cite{CapTri}. Recent review of
theoretical models and state-of-art in theory could be found in,
e.g., \cite{Cape2,CapBok}. So far, two main groups of theoretical
models of purely quantum open systems violating the second law
existed: Those with quantum reaction channels opening or closing
in accordance with the instantaneous state of the system
(reminding of the Maxwell demon \cite{Maxw} closing and opening
gate in a wall separating two compartments with a classical gas),
and those where a specific type of interference of different
quantum reaction channels exists \cite{Cape2}. In particular this
type of systems is relevant as the contradiction with the second
law treated in \cite{Cape2} is not only mathematically well
justified but can be given even a very simple physical
interpretation based on otherwise experimentally well established
facts: Exciton diffusion bearing energy and going (as always
diffusion does) in the direction of decreasing exciton
concentration.

Recently, as a theoretical response to another positively tested
experimental system \cite{Shee1}, a next model of still another
type has been suggested \cite{CapShe} that also allows rigorous
solution (exact within a scaling theory) fully confirming
experimental doubts about universal validity of the second law.
Detailed discussion and solution of a modified (and nearer to
reality) version of the model is the subject of the present paper.
Here, however, we are already able to support the conclusions by
both scaling and non-scaling arguments. What is perhaps universal
for all the models challenging thermodynamics is that the system
in question must be, during its activity, outside the canonical
state \cite{AllNie}\footnote{A comment might be relevant in this
connection that outside equilibrium, there is no reason why the
thermodynamic entropy (characterizing individual macroscopic body)
and the Shannon - von Neumann entropy (characterizing ensemble,
i.e. bearing another physical information) should be identical.
For validity of the second law in Nature, the former entropy is
relevant while only the latter is usually involved in analytical
`proofs' of the second law starting from first principles.} .
Mechanisms how to achieve and maintain that might be, of course,
different. One should understand, however, that this condition is
perhaps necessary but by far not sufficient. For other paradoxical
systems that could be also classified as above, showing how
physics beyond the second law is still little understood see,
e.g., \cite{Shepar,Gord}.

\section{Theoretical model}

Physicists are more conservative than mathematicians. For the
latter, one unquestioned proof is standardly enough to accept a
new statement, theorem etc. Physicist on the other hand require
usually more arguments, especially when old truths or dogmas are
challenged. That is why we should like to stress already here
that:
\begin{itemize} \item The model treated here is by no means
unique. It is just a next one in a long series of models behaving,
according to a strict microscopic theory, in a way incompatible
with standard thermodynamics. Among those, it is exceptional just
in the sense of modelling an experimental system tested
positively, from the point of view of violability of the second
law, already a long time ago.
\item Except for model assumptions, we shall use no approximate
steps in our reasoning in the main text here. This is to be
stressed in order to avoid misunderstandings.
\item Mathematics we are going to use in the main text
is that one by Davies \cite{Davi1,Davi2} which forms, for one
specific choice of the scaling parameter, mathematical basis of
the weak-coupling kinetic relaxation theory confirming, for the
weak-coupling limit, validity of the second law. (Another
possibility of deriving basic kinetic equations (\ref{Lioeq})
below may be connected with the Tokuyama - Mori identity stemming
from the Heisenberg equations of motion for quantum operators -
see, for another model, e.g. \cite{CapBok}.) We fully rely upon
the Davies mathematics and use, for the chosen model (i.e. a
specific case) no additional approximations. Hence, rigour of our
approach below is that one of the general Davies theory. We only
deviate from a subsequent standard application of general Davies
theorems to the weak coupling situation by just another, but
equally admissable and physically motivated, choice of the scaling
parameter. This choice makes the theory physically applicable also
beyond the weak coupling limit.
\item In order to convince sceptics
concerning physical applicability of the scaling ideas, we also
present completely independent mathematical non-scaling arguments.
Those may be found in the Appendix and are important in particular
in the relevant long-time limit. \footnote{Doubts recently
appeared about ability of the Davies scaling theory to yield
reliable long-time predictions on the relaxation \cite{Novo}. The
exactly solvable model considered was, however, bilinear in
creation and annihilation operators. Such models, like that of
harmonic phonons, are known on the other hand to be unable to
describe diffusion regime underlying, e.g., the effect discussed
in \cite{Cape2}. Thus, \cite{Novo} should be taken only as a
warning against thoughtless application of formal conclusions of
the Davies theory to, e.g., diffusion flows where no such flows
exist. Here, it means another formal argument for supporting
conclusions of the (otherwise broadly tested) Davies theory in our
main text by independent arguments in the Appendix.}
\end{itemize} As we have numerically verified and as it is also
argued below, our results well coincide with those of the
weak-coupling theories in the overlap region with the
weak-coupling regime. We thus have no doubts on the validity of
the second law there. However, beyond the weak coupling regime,
our results become appreciably different from standard ones.
Physically, we have reasons supported by arguments to understand
that: This is a deviation from the canonical state of the system
caused by its non-negligible coupling to the bath. As far as the
underlying mathematics is concerned, it is general (valid for any
choice of the scaling parameter) and cannot be consequently
sometimes correct and sometimes not. It can be either correct or
not in general; no other alternative exists. The first alternative
provides solid basis for arguments in favour of correctness of our
approach while the second one (that was even never suggested or
indicated) deprives even the weak coupling relaxation theory of
its mathematical foundation. Rejecting these alternatives would
\begin{itemize}
\item either mean to question physical as well as mathematical
principles (including the Liouville equation) on which all the
existing renown of description of kinetic phenomena via
corresponding kinetic equations (depending on the regime in
question) relies,
\item or to admit that the very principles of quantum mechanics of
the open system would have to be complemented by, e.g., some
additional requirements not admitting models of the type
investigated here.
\end{itemize} In view of existing experimental evidence in favour
of the quantum theory as well as because of experimental results
indicating violations of the second law in experiment
\cite{Shee1,SheMea,Niku2} (see also \cite{Niku3} for theoretical
interpretation of the experimental results reported in
\cite{Niku2} as well as previously \cite{NikZhi}), both the latter
possibilities seem, in our opinion, rather unlikely. In any case,
the conclusions suggest that there is at least something in
physics beyond the second law what is at present still
insufficiently understood.

Our system, in accordance with the experimental plasma system
\cite{Shee1}, is assumed to consist of three sites, designated as
1, 2, and 3. The reader is referred to \cite{Shee1} or
\cite{CapShe} if he/she is interested in the motivation for
construction of the model. The latter is, as compared to
\cite{CapShe}, only slightly modified here so that it now better
corresponds to the experimental system of \cite{Shee1}. Shortly,
the above sites correspond to walls of the plasma container,
plasma, and the probe. Hamiltonian of the system reads
\be
H_S=\sum_{j=2}^3\epsilon_ja_j^{\dagger}a_j+J(a_1^{\dagger}a_2+a_2^{
\dagger}a_1)+K(a_2^{\dagger}a_3+a_3^{\dagger}a_2) \label{H_S}, \ee
where the zero of the energy is taken to be at the walls (site 1).
Though it is not in principle important, we assume here
$\epsilon_3>\epsilon_2>0$. This corresponds to the experimental
situation \cite{Shee1}. We assume only one electron in the system
that is elastically transferred between sites 1 and 2, and
simultaneously between sites 2 and 3. This is in accordance with,
e.g., the standard theory of the Richardson-Dushman thermal
emission (for the 1 - 2, i.e. wall - plasma transfers) that is
based on the idea of prevailingly elastic transfer upon, e.g.,
electron leaving surface of solids.

The load between the probe and walls in the Sheehan's experimental
set-up \cite{Shee1} is the location at which the electron can
inelastically scatter. This means, in our case, phonon-assisted
$3\leftrightarrow1$ transitions. The phonons involved are assumed
to be those of the load, here designated as bath II. In addition
to that, we assume another bath, say bath I, formed by phonons
(physically, those from the walls) interacting site-locally with
the electron located on site 1. This means that Hamiltonian $H_B$
of the bath of our model reads as
\[ H_B=H_B^I+H_B^{II}, \]
\be H_B^I=\sum_{\kappa}\hbar\omega_{\kappa}b_{\kappa}^{\dagger}
b_{\kappa}, \quad H_B^{II}=\sum_{\kappa}\hbar\omega_{\kappa}B_{
\kappa}^{\dagger} B_{\kappa}. \label{H_B} \ee
The electron-bath coupling, $H_{S-B}$, is given by
\[H_{S-B}=H_{S-B}^I+H_{S-B}^{II},\]
\[H_{S-B}^I=\frac{1}{N}\sum_{\kappa_1\neq\kappa_2}\hbar\sqrt{\omega_{
\kappa_1}\omega_{\kappa_1}}g_{\kappa_1,\kappa_2}a_1^{\dagger}a_1
(b_{\kappa_1}+b_{\kappa_1}^{\dagger})(b_{\kappa_2}+b_{\kappa_2}^{
\dagger}),\] \be H_{S-B}^{II}=\frac{1}{\sqrt{N}}\sum_{\kappa}\hbar
\omega_{\kappa}G_{ \kappa}(a_3^{\dagger}a_1+a_1^{\dagger}a_3)
(b_{\kappa}+b_{\kappa}^{\dagger}). \label{H_S-B} \ee (Contingent
terms with $\kappa_1=\kappa_2$ in $H_{S-B}^I$ could be turned
below to just a temperature-dependent renormalization of
site-energy $\epsilon_1$ of site 1 that we set zero here.) Here
the anti- or commutational relations between creation and
annihilation operators for electrons and phonons are as usual.
Also $N$ designates the number of phonon modes; here it is
understood that $N\rightarrow+\infty$. Notice that $H_{S-B}^{I}$
is quadratic in the phonon operators; this is certainly admissable
but contrasts with both tradition and the treatment of phonon
operators in $H_{S-B}^{II}$. This non-standard assumption is
employed to preserve the finite dephasing and local electron
heating after the Davies scaling procedure (preserving formally
just the lowest-order effects in the scaling parameter) which now
follows. Otherwise, we would have to involve higher-order effects
in treating the dephasing what could make the theory and the final
statements, in eyes of a sceptic reader, rather ambiguous. In any
case, we have a freedom to choose the model as above. Concerning
$H_{S-B}^I$, one should also notice that the on-site dephasing is
here, in contrast to the model from \cite{CapShe}, on site 1. This
corresponds, in the experimentally tested system of \cite{Shee1},
to electron heating inside the walls of the plasma container.

One should first of all realize that the phonon-assisted transfers
$3\leftrightarrow1$ as provided by $H_{S-B}^{II}$ enable the
electron to move in a circle
$1\rightarrow2\rightarrow3\rightarrow1$, or {\em vice versa}
$1\leftarrow2\leftarrow3\leftarrow1$. These two circular motions
cannot, however, compensate each other. This is the first of two
basic physical observations on which the present model relies. The
point is that all the inter-site transfers involved are elastic,
i.e. symmetric, except for the $3\leftrightarrow1$ one. Such
elastic transfers lead to a tendency of equilibration of site
occupation probabilities. For instance, assume for a while that we
had only a dimer composed of sites 1 and 2, with the coherent
(i.e. elastic) hopping term $J(a_1^{\dagger}a_2+a_2^{\dagger}a_1)$
decoupled from any bath. The Hamiltonian reads then
$H_S=\epsilon_2^{\dagger}a_2 +J(a_1^{\dagger}a_2+H.C.)$ as above.
Stationarity of the solution we are interested in implies
$\rho_{12}=\rho_{21}$. Contingently nonzero values of these
site-off-diagonal elements are connected with a 1-2 bonding. If we
add a mechanism breaking such bonds (but, for simplicity, causing
no additional $1\leftrightarrow2$ transfer), magnitude of
$\rho_{12}=\rho_{21}$ would become suppressed below their maximum
value given by positive semi-definiteness of $\rho$. Concerning
the site-diagonal elements $\rho_{jj}$, their stationary values
can be investigated by generalized master equations where memory
functions determine the $1\leftrightarrow2$ balance. These memory
functions (and also their time integrals whose ratio determines
$\rho_{11}(+\infty)/\rho_{22}(+\infty)$) contain two channels
\cite{CapZP}, sometimes interpreted as phonon- (or bath-) assisted
and quasicoherent one \cite{Kenkre}. The latter channel is, in
contrast to the former one, symmetric what is the reason of
generally comparable stationary values of $\rho_{11}(+\infty)$ and
$\rho_{22}(+\infty)$. Full equilibration
$\rho_{11}(+\infty)=\rho_{22}(+\infty)$ then follows from our
mathematics below upon full ignoring site 3, as a consequence of
elastic character of the $1\leftrightarrow2$ transitions. The same
applies for dimer 2 and 3 once it is separated from the rest of
the system. These facts will be useful below. With that, it is
then easy to show that the equality of populations
$\rho_{11}\approx\rho_{33}$ stemming from above reasoning based on
the elastic character of the $J$- and $K$-induced elastic
transfers cannot comply with equilibration of the inelastic
phonon-assisted transitions $3\leftrightarrow1$ leading
$\rho_{11}>\rho_{33}$ or possibly even to $\rho_{11}\gg\rho_{33}$.
This fact will be useful to understand the results obtained below.
A word of warning is, however, worth already here: What is here
now being explained are still just heuristic arguments supported
by previous investigations that explain our motivation; true
rigorous mathematics comes only below. As it follows from general
arguments above as well as the mathematics of the next sections,
we do understand why, e.g., the detailed balance conditions could
in our system become violated. This is because these conditions do
not apply to uphill or downhill transfers caused by elastic
mechanisms. We, however, do not raise the question about
violations or preserving these conditions here. Our rigorous
mathematics below avoids such statements and formulations, and
leads directly to the required results and effects investigated
here.

For that, let us return to our model (\ref{H_S}-\ref{H_S-B}).
Between sites 1 and 3, there is an imbalance mechanism owing to
spontaneous processes allowed by  $H_{S-B}^{II}$ that prefers,
because of assumed $\epsilon_3>0$, the $3\rightarrow1$ transitions
to $1\rightarrow3$ ones. This is the imbalance (and the only
imbalance existing in our system) that makes domination of the
circular motion $1\rightarrow2 \rightarrow3\rightarrow1$ over the
$1\leftarrow2\leftarrow3 \leftarrow1$ one in fact possible. Owing
to the phonon-assisted (prevailingly down-hill, i.e.
$3\rightarrow1$) character of the $3\leftrightarrow1$ transfer,
this implies heat transfer to bath II. (Each transfer act
$3\rightarrow1$ is connected, because of the energy conservation
law, with emission of a phonon quantum into bath II. Similarly for
the back transfer $3\leftarrow1$ and phonon absorption. If the
former transfers prevail, we get the net heat flow to bath II.)
The question is, however, where this heat could come from. The
only possible answer is that it is from bath I. Really, dephasing
at site 1 means nothing but a continuous emission and absorption
of phonons from bath I that can provide the necessary energy
transferred by the electron whose energy is not sharp.
Simultaneously, this dephasing can break phase relations between
amplitude of finding the electron at site 1 and those elsewhere,
i.e. it breaks the corresponding covalent-type of bonds. Without
the sufficiently strong dephasing, the electron distribution in
the system (as prescribed by, e.g., the canonical density matrix
to which the density matrix usually tends within weak-coupling
theories considering coupling to bath as infinitesimally weak)
would really contain such {\em and fully developed} $J$- and
$K$-induced bonds. Thus, it would be stiff enough in the respect
that irrespective of the above imbalance, no electron circular
motion would finally appear. One can easily verify that by
calculating, e.g., the electron flow between any two sites. We
always get (by the way, in a correspondence with standard physical
reasoning) zero mean flow in the canonical (i.e. zeroth-order in
the coupling to the bath) state of the system. From this point of
view, lack of on-site amplitude dephasing (or, in other words,
that of partial violations of such covalent bonds re-appearing
immediately once the system coupling to the bath is re-introduced
as a source of corrections to the canonical form of its density
matrix) is one of the greatest deficiencies of the weak-coupling
kinetic approaches leading to such canonical distributions. For
illustration, notice that site off-diagonal elements of the
density matrix become, in the weak-coupling kinetic theories,
asymptotically independent of the strength of the site-local
coupling to the bath even when this type of the coupling causes
bath-induced (and on different sites uncorrelated) fluctuations of
site energies (see, e.g., formula (28) of \cite{CaToMo}). So, even
when the model does describe it, its weak-coupling kinetic
relaxation formalism is in principle unable to describe the
dephasing and breaking of the bonds. Hence, in our case here, we
must definitely go beyond the weak coupling kinetic theory in
especially the dephasing rate at site 1. This is the second basic
and, perhaps, the most important observation connected with the
model and the effect we should like to describe here. (One could
also pose a question why we should be so keen to describe and
include such potentially weak corrections to, e.g. violations of
covalent $1-2$ and $2-3$ bonds here. The point is that these
violations provide perhaps just small corrections to the canonical
density matrix of the system but, simultaneously, these
corrections are the very source of the effect we are interested
in.) One should add that in our approach involving other than the
weak coupling approach below, we also take the limit of the
infinitesimal coupling to bath, i.e. infinitesimal dephasing. On
the other hand, we simultaneously scale also the hopping
(transfer) integrals what makes the ratio of the in-phasing and
dephasing constant. This, in contract to the standard weak
coupling scaling, corresponds to reality and allows the above bond
breaking. (In the Sheehan experimental plasma system \cite{Shee1},
no such $1-2$ or $2-3$ covalent bonding exists). Thus we have a
hope, and really do obtain below the effect expected. So far, of
course, all these ideas provide us with at most a physical
background of the model, its mathematical treatment, and physics
beyond. So, let us now have a look at how these ideas work within
a rigorous theory.

\section{Davies Scaling and Kinetic Equations}

The scaling procedure we use is based on Davies
\cite{Davi1,Davi2}. (For independent non-scaling arguments fully
supporting results obtained here see the Appendix.) We, however,
extend our treatment beyond standard weak coupling theory in that
we scale not only time and $H_{S-B}$, but also the transfer
(overlap, hopping, or resonance) integrals $J$ and $K$, setting
\cite{CapBar}
\be t=t'/\lambda^2,\quad H_{S-B}\propto\lambda,\quad J\propto
\lambda^2, \quad K\propto\lambda^2 \label{scale} \ee ($t'$ playing
the role of a new rescaled time.) As usual, we then project off
the bath and let $\lambda\rightarrow0$. Such a physical regime
where intersite hopping (transfer) integrals determining rates of
bath-free transfers inside system get comparable with rates of
bath-assisted processes (transfers) can be definitely {\em not}
that of the weak coupling but rather that of the intermediate or,
in a sense, even contingently strong coupling to the bath.
Technically, though the mathematics used is completely that by
Davies \cite{Davi1}, we proceed simultaneously according to
\cite{CapBar} where the relevant formulae are rewritten in a
physically understandable form.

The Davies formalism starts by writing total Hamiltonian
\be H=H_S+H_B+H_{S-B}\ee
in form
\be H=H_0+\lambda H_1. \label{split} \ee
Here, one should add that $\lambda H_1\propto\lambda$ but that does not
exclude the possibility that $\lambda H_1$ includes also higher orders
in $\lambda$ ($\lambda^2$ if (\ref{scale}) is accepted). Those who do not
like this way of thinking could replace conditions $J\propto \lambda^2$,
$K\propto\lambda^2$ in (\ref{scale}) by $J\propto\lambda$, $K\propto\lambda$
and proceed as below. The final result is the same.

Next, introduce superoperators
\be {\cal L}_0=\frac{1}{\hbar}[H_0,\ldots],\quad {\cal
L}_1=\frac{1}{\hbar}[\lambda H_1,\ldots]\propto\lambda.
\label{Livspl} \ee Finally, be
\be {\cal P}\ldots=\rho^B\otimes{\rm Tr}_B(\ldots) \label{ArgKel} \ee
(with ${\rm Tr}_B\rho^B=1$ implying ${\cal P}^2={\cal P}$) the
Argyres-Kelley projector (projection superoperator) in the Liouville space
of operators that act in the Hilbert space of the system and bath.
Then the message of Davies (see Eq. (1.19) of \cite{Davi1} or Eq. (14) of
\cite{CapBar}) is
\[ \lim_{\lambda\rightarrow0}\sup_{0\le\lambda^2t\le a}||\rho(t)-{\rm
e}^{-i({\cal L}_0+\langle{\cal L}_1\rangle+i\lambda^2{\cal
K})t}\rho(0)||=0,\]
\[\lambda^2 {\cal K}...=\int_0^{+\infty}dx\,Tr_B\ ({\rm
e}^{i{\cal L}_0x}(-i{\cal L}_1){\rm e}^{-i{\cal L}_0x}(1-{\cal
P})(-i{\cal L}_1)(\rho^B\otimes...)),\] \be \langle{\cal
L}_1\rangle...=Tr_B(\rho^B\otimes{\cal L}_1...). \label{Davres}
\ee (Finite constant $a$ is here arbitrary.) Here $\rho(t)={\rm
Tr}_B\rho^{S+B}(t)$ is the density matrix of the complex `system +
bath' with its time-development determined from the exact
Liouville equation $i\frac{d}{dt}\rho^{S+B}(t)=({\cal L}_0+{\cal
L}_1)\rho^{S+B}(t)$. The assumptions used were in particular
\begin{itemize}
\item that the density matrix of the system and bath
$\rho^{S+B}(t)$ is initially separable, i.e. that
\be \rho^{S+B}(0)=\rho^B\otimes\rho(0), \label{inicond} \ee and
\item that ${\cal PL}_0={\cal L}_0{\cal P}$. This condition can be,
however, well fulfilled as far as, e.g.,
$\rho^B=f(H_B^I,H_B^{II})$, $[H_B^I,H_B^{II}]=0$. This is in
particular in our case because we are forced to assume
\be \rho^B=\frac{\exp(-\beta_IH_B^I-\beta_{II}H_B^{II})}{{\rm
Tr}_B \exp(-\beta_IH_B^I-\beta_{II}H_B^{II})}  \label{InConBa} \ee
to be able to introduce properly the initial temperatures of baths
$I$ and $II$ separately.
\end{itemize} Suspicious reader with potential objections about
applicability of the scaling theories to, in particular, the
long-time relaxation phenomena is again referred to the Appendix,
in particular its concluding remarks. Here, we should just like to
add one comment: Situations are known when (\ref{Davres}) cannot
be used for finite $\lambda's$ (as in Nature). Perhaps the
simplest model of this type is that of a single particle on a
periodic chain where the system is artificially introduced via a
few chosen sites (with the particle or without it), with the rest
being the (formal) bath and the system-bath coupling (bringing the
particle to or out of the system). Then matrix elements of
$\rho(t)$ between sites of the system decay algebraically
\cite{Kenkre} while ${\rm e}^{-i({\cal L}_0+\langle{\cal
L}_1\rangle+i\lambda^2{\cal K})t}\rho(0)$ would indicate
exponentially dominated decay. The model by Novotn\'{y}
\cite{Novo} is also of this type. With our real bath known to
yield a well-defined relaxation, we shall for the sake of brevity
here postpone discussion of such singular cases with artificially
introduced baths to another publication.

Meaning of the mathematically exact statement in the first row of
(\ref{Davres}) is that time development of $\rho(t)$ as prescribed by the
exact Liouville equation for the density matrix $\rho^{S+B}(t)$ of the
complex `system + bath' is not discernable, in the scaling limit
$\lambda\rightarrow0$, from that one dictated by the kinetic equation for
the density matrix $\rho(t)$ of just the system
\be i\frac{d}{dt}\rho(t)=({\cal L}_0+\langle{\cal L}_1\rangle+i\lambda^2{\cal
K})\rho(t). \label{Lioeq} \ee
This general and exact result should now be specified according to the
choice of $H_0$ and $\lambda H_1$ in (\ref{split}).

Two main possibilities exist. \begin{itemize} \item Either we accept so
called weak-coupling scaling according to van Hove (and often
automatically accepted in general situations)
\be t=t'/\lambda^2,\quad  H_{S-B}\propto\lambda,\quad J={\rm
const},\quad K={\rm const} \label{scaleVH} \ee (again with
$\lambda\rightarrow0$) which would correspond to the choice
\be H_0=H_S+H_B, \quad \lambda H_1=H_{S-B}. \label{WCchoi} \ee
Then $\lambda^2{\cal K}$ in (\ref{Lioeq}) is nothing but the
weak-coupling relaxation superoperator and (\ref{Lioeq}) reduces
to the Redfield equation (before using the Redfield secular
approximation) \cite{Redf1,Redf2,MahWeb}. The relaxation is then
practically to the canonical state of the system. \item Or we
assume (\ref{scale}) what means to identify
\be H_0=H_S|_{J=K=0}+H_B, \quad \lambda
H_1=H_{S-B}+H_S|_{J\neq0,\,K\neq0}-H_S|_{J=K=0}. \label{CapBarid}\ee
Then several things have to be realized:
\begin{itemize} \item Though the relaxation superoperator $\lambda^2{\cal
K}$ in (\ref{Lioeq}) [as defined in (\ref{Davres})] involves
formally also higher orders in $\lambda$, application of the rule
${\rm Tr}_B([a_m^{\dagger}a_n,\ldots])=[a_m^{\dagger}a_n,{\rm
Tr}_B(\ldots)]$ yields that in fact just second order terms in
$\lambda$ survive.
\item Because now ${\cal L}_0=\frac{1}{\hbar}[H_S|_{J=K=0}+H_B,\ldots]$
and $H_S|_{J=K=0}$ is site-diagonal, the relaxation is {\em not any more}
(like in case of the weak-coupling choice (\ref{WCchoi})) among eigenstates
of $H_S$ but, instead, among those (site-local eigenstates) of
$H_S|_{J=K=0}$. On the other hand, in (\ref{Lioeq}), terms ${\cal
L}_0+\langle{\cal L}_1\rangle$ reproduce
$\frac{1}{\hbar}[H_S,\ldots]\equiv\frac{1}{\hbar}[H_S|_{J\neq0\neq K}
,\ldots]$ what is a free propagation among eigenstates of the full
(site-off-diagonal) $H_S$. This competition between site-local and site
off-local tendencies of the time development is what makes the dynamics
much richer than in the weak-coupling case. \item The fact that
$\lambda^2{\cal K}$ describes relaxation in the site-local basis is
{\em not} owing to neglecting anything or any type of approximation. It is
owing to choice of another regime; in our case that one in which $J$- and
$K$-induced processes become at most comparable with those caused by
the system interaction with the bath.
\end{itemize} \end{itemize}

The physical argument in favour of the form of the relaxation
superoperator corresponding to the choice (\ref{CapBarid}) is that
we are interested in the regime in which the bath-assisted
processes inside the system are at least comparable with, or even
dominating over the internal transfer processes inside the system
caused, in our case, by the $J$- and $K$-dependent hopping terms
in $H_S$ (\ref{H_S}). Once we realize that the weak-coupling
theory {\em presumes}, in the sense of (\ref{scaleVH}), the
system-bath coupling to be infinitesimal, i.e. infinitely times
weaker than all other relevant competing transfer and relaxation
mechanisms, this excludes scaling (\ref{scaleVH}), i.e. the choice
(\ref{WCchoi}), in the regime considered here. On the other hand,
it allows to use (\ref{scale}), i.e. the choice (\ref{CapBarid}).
That is why we shall below stick to this alternative. So, we use
${\cal K}$ from (\ref{Davres}) with (\ref{CapBarid}) for our model
(\ref{H_S}-\ref{H_S-B}). This means the Redfield form of ${\cal
K}$ in the localized basis as a consequence of another (than the
weak-coupling) physical regime, i.e. also correspondingly another
form of identification of perturbation. So, this form of the
Redfield tensor is definitely {\em not}  consequence of any
additional approximation applied to the Redfield form of the
relaxation superoperator in the weak-coupling regime (for
discussion of such an approximation in the weak-coupling regime
see \cite{Kleine}). After some straightforward algebra,
(\ref{Lioeq}) then turns in the site representation to
\[ i \hbar \frac{d
\rho}{dt}\equiv i\hbar\frac{d}{dt}\left(\begin{array}{c}
\rho_{11}\\ \rho_{22} \\ \rho_{33} \\ \rho_{12} \\ \rho_{21}
\\ \rho_{13} \\ \rho_{31} \\ \rho_{23} \\ \rho_{32}
\end{array}\right)=\left(\begin{array}{ccccccccc} ~ & ~ & ~
& ~ & ~ & . & ~ & ~ & ~\\ ~ & ~ & ~ & ~ & ~ & . & ~ & ~ & ~ \\
~ & ~ & {\cal A} & ~ & ~ & . & ~ & {\cal B} & ~\\ ~ & ~ & ~ &
~ & ~ & . & ~ & ~ & ~ \\ ~ & ~ & ~ & ~ & ~ & . & ~ & ~ & ~ \\
~^{^{\cdot}} & ~^{^{\cdot}} & ~^{^{\cdot}} & ~^{^{\cdot}} & ~^{^{
\cdot}} &.^{^{\cdot}} & ~^{^{\cdot}} & ~^{^{\cdot}} & ~^{^{\cdot}}
\\ ~ & ~ & ~ & ~ & ~ & . & ~ & ~ & ~ \\ ~ & ~ & {\cal B}^T & ~
& ~ & . & ~ & {\cal C} & ~ \\ ~ & ~ & ~ & ~ & ~ & . & ~ & ~ & ~
\end{array}\right)\cdot \left(\begin{array}{c} \rho_{11}\\ \rho_{22}
\\ \rho_{33}\\ \rho_{12}\\ \rho_{21}\\ \rho_{13}\\ \rho_{31}\\
\rho_{23}\\ \rho_{32} \end{array}\right).\] \be \label{Liou} \ee
The sub-matrices ${\cal A}$, ${\cal B}$, ${\cal C}$ (${\cal B}^{T}$ is the
transpose of ${\cal B}$) are given as
\[{\cal A}=\left(\begin{array}{ccccc} -i\hbar\Gamma_{\uparrow} & 0 &
i\hbar\Gamma_{\downarrow} & -J & J \\ 0 & 0 & 0 & J & -J \\
i\hbar\Gamma_{\uparrow} & 0 & -i\hbar\Gamma_{\downarrow} & 0 & 0 \\
-J & J & 0 & -2i\hbar\Gamma-\frac{i\hbar}{2}\Gamma_{\uparrow}-
\epsilon_2 & 0\\ J & -J & 0 & 0 & -2i\hbar\Gamma-\frac{i\hbar}{
2}\Gamma_{\uparrow}+\epsilon_2 \end{array}\right), \]
\[{\cal B}=\left(\begin{array}{cccc} 0 & 0 & 0 & 0 \\ 0 & 0 & -K
& K \\ 0 & 0 & K & -K \\ -K & 0 & 0 & 0 \\ 0 & K & 0 & 0 \end{array}\right), \]
\[ {\cal C}=\left(\begin{array}{cccc}-2i\hbar\Gamma-\frac{i\hbar}{2}
(\Gamma_{\uparrow}+\Gamma_{\downarrow})-\epsilon_3 & \frac{i\hbar}{
2}(\Gamma_{\uparrow}+\Gamma_{\downarrow}) & J & 0 \\ \frac{i\hbar}{2}
(\Gamma_{\uparrow}+\Gamma_{\downarrow}) &-2i\hbar\Gamma-\frac{i\hbar}{2}(
\Gamma_{\uparrow}+\Gamma_{\downarrow})+\epsilon_3 & 0 & -J \\ J &
0 & -\frac{i\hbar}{2}\Gamma_{\downarrow}+\epsilon_2-
\epsilon_3 & 0\\ 0 & -J & 0 & -\frac{i\hbar}{2}
\Gamma_{\downarrow}-\epsilon_2+\epsilon_3\end{array}\right). \]
\be \label{matr} \ee
Here, we have used the notation
\[\Gamma_{\uparrow}=\frac{2\pi}{\hbar}\frac{1}{N}\sum_{\kappa}
|\hbar\omega_{\kappa}|^2|g_{\kappa}|^2n_B(\beta_{II},\hbar\omega_{
\kappa})\delta(\hbar\omega_{\kappa}-\epsilon_3), \]
\[\Gamma_{\downarrow}\equiv\frac{2\pi}{\hbar}\frac{1}{N}\sum_{\kappa}
|\hbar\omega_{\kappa}|^2|g_{\kappa}|^2[1+n_B(\beta_{II},\hbar\omega_{
\kappa})]\delta(\hbar\omega_{\kappa}-\epsilon_3)=\Gamma_{\uparrow}
\cdot{\rm e}^{\beta_{II}\epsilon_3}, \]
\[ 2\Gamma=\frac{2\pi}{\hbar}\frac{1}{N^2}\sum_{\kappa_1,\kappa_2}
|g_{\kappa_1,\kappa_2}|^2(\hbar\omega_{\kappa_1}\omega_{\kappa_2})^2
n_B(\beta_I,\hbar\omega_{\kappa_1})[1+n_B(\beta_I,\hbar\omega_{\kappa_2})]
\delta(\hbar\omega_{\kappa_1}-\hbar\omega_{\kappa_2}),\] \be
n_B(\beta,z)=\frac{1}{{\rm e}^{\beta z}-1}, \label{Lionot} \ee
where $T_{I (II)} =1/(k_B\beta_{I (II)})$ are the initial
temperatures of Baths I and II;  $n_B(\beta, z)$ is the
Bose-Einstein phonon distribution function. $\Gamma_{\uparrow}$
and $\Gamma_{\downarrow}$ are the Golden Rule formulae for
transfer rates $1\rightarrow3$ and $3\rightarrow1$. Note that
$\Gamma_{\uparrow}$ and $\Gamma_{\downarrow}$ are different solely
in that the latter involves a $1 + n_{B}$ term, whereas the former
has only $n_{B}$.  Physically, this corresponds to
$\Gamma_{\uparrow}$ involving only bath-assisted stimulated
up-hill transitions (absorption), whereas $\Gamma_{ \downarrow}$
involves both bath-assisted spontaneous and bath-assisted
stimulated down-hill transitions (emission). Finally, $2\Gamma$
determines the rate of dephasing arising from local
electron-energy fluctuations from Bath I, and also the rate of
electron heating in the walls (site 1).

A few comments are worth mentioning already here. First, notice that
temperature $T_I$ of bath I enters (\ref{Liou}) only via the dephasing
(and simultaneously heating) rate $2\Gamma$. This rate depends, however,
also on strength and details of the electron coupling to bath I.
Thus, moderate changes of $T_I$ may be well compensated by those of the
coupling and {\em vice versa}. As there are no abrupt qualitative changes
expected with moderate changes of the coupling, only continuous changes
of, e.g., the electron $1\rightarrow2\rightarrow3$ flow are expected when
(the initial) temperature $T_I$ of bath I sinks below that (i.e. $T_{II}$)
of bath II. This is important for interpretation of the result to be
obtained below. Finally, concerning (the initial) temperature $T_{II}$
of bath II: We shall assume here the inequality
\be k_BT_{II}\stackrel{<}{\sim}\epsilon_3. \label{Tnon1}\ee The
opposite inequality would imply high-temperature regime in which
the spontaneous $3\rightarrow1$ processes would become negligible
with respect to stimulated ones. So, asymptotically,
$\Gamma_{\uparrow}/ \Gamma_{\downarrow}$ would turn to unity and
the driving force in the circle
$1\rightarrow2\rightarrow3\rightarrow1$ would disappear implying
disappearance of the electron flow. This means that also the
contradiction with the second law we aim at would disappear, in a
full correspondence with the Bohr correspondence principle and the
Martynov proof of validity of the second law in classical
statistical mechanics \cite{Mart}. On the other hand, the
low-temperature limitation (\ref{Tnon1}) is not severe. With,
e.g., $\epsilon_3\approx 1$eV, temperatures $T_{II}$ appreciably
higher than room temperatures are viable.

\section{Analytical solution and steady-state heat-flow}

Let us henceforth investigate the stationary situation. Then the
left hand side of (\ref{Liou}) equals zero so that we have a
homogeneous set of 9 linear algebraic equations for the stationary
values of the electron density matrix. The matrix rank is,
however, only 8 since the sum of its first three rows is zero.
Thus, the set can (and must) be complemented by the normalization
condition
\be \rho_{11}+\rho_{22}+\rho_{33}=1. \label{norm} \ee This
provides us with a complete inhomogeneous set of 9 linear
algebraic equations for 9 elements of the particle density matrix.
The site-diagonal matrix elements $\rho_{jj}$ give the
probabilities of finding the electron at site $j$. Full algebraic
solution of this set of equations is possible but unwieldy. That
is why the set will be solved numerically below. First, however,
we shall analytically prove that there is always, for nonzero
temperatures $T_I$ and $T_{II}$, a positive electron flow
$1\rightarrow2\rightarrow3\rightarrow1$ implying, for
$T_I<T_{II}$, violation of the second law of thermodynamics in its
Clausius formulation. The proof is made by logical contradiction.
We stress from the outset that the mathematical derivation of our
starting equations (\ref{Liou}), including scaling, involves no
approximation; therefore, aside from the model assumptions, our
treatment here is fully rigorous, in the full mathematical meaning
of the word.

Cyclic mean electron flow in the system (taken as positive in the
direction $1\rightarrow2\rightarrow3\rightarrow1$) can be written,
on grounds of physical meaning of $\Gamma_{\uparrow}$ and
$\Gamma_{\downarrow}$ in (\ref{Lionot}), as
\be {\cal
J}=\Gamma_{\downarrow}\rho_{33}-\Gamma_{\uparrow}\rho_{11}.
\label{flo1} \ee
From the first and second equations of
(\ref{Liou}) one also has
\be {\cal
J}=\frac{i}{\hbar}J(\rho_{21}-\rho_{12})=\frac{i}{\hbar}K
(\rho_{32}-\rho_{23}). \label{flo2} \ee These formulae can be also
easily derived from elementary quantum mechanics. Assume now that
no heat flows directly from bath I to bath II and {\em vice
versa}. So, only the electron-mediated heat flow from I to II may
appear. Since the $3\leftrightarrow1$ phonon-assisted transitions
are associated with influx or efflux of energy (heat) to or from
Bath II (proportionally to the magnitude of $\epsilon_{3}$), the
total mean heat flow from Bath I to Bath II may be written as
\be {\cal Q}=\epsilon_3{\cal J}. \label{Qflo} \ee
This is the main quantity we are interested in.

Let as now {\em assume}, in accord with our strategy of proof by
contradiction, that there is no heat flow between the baths, i.e.
\be {\cal Q}=0. \label{Ass1} \ee
Since  $\epsilon_3>0$, this implies, via (\ref{flo1}) and (\ref{Qflo}),
that
\be \Gamma_{\downarrow}\rho_{33}-\Gamma_{\uparrow}\rho_{11}=0.
\label{Ass2} \ee The reader could easily recognize that
(\ref{Ass2}) is nothing but a detailed balance condition for
inelastic phonon-assisted $1\leftrightarrow3$ direct transitions.
Because of (\ref{flo2}) this also implies that
\be \rho_{21}=\rho_{12},\quad \rho_{32}=\rho_{23}. \label{Ass3} \ee
Now, summing the forth and fifth equations of (\ref{Liou}) with zero left
hand side, we get
\be 0=\frac{K}{\hbar}(\rho_{31}-\rho_{13})+(-2i\Gamma-\frac{i}{2}
\Gamma_{\uparrow})(\rho_{12}+\rho_{21})+\frac{\epsilon_2}{\hbar}
(\rho_{21}-\rho_{12}). \label{eqA} \ee Similarly, from the sixth
and seventh equation, and also from the eighth and ninth equation
of (\ref{Liou}) (always with zero left hand side), we get
\be
0=\frac{K}{\hbar}(\rho_{21}-\rho_{12})-2i\Gamma(\rho_{13}+\rho_{31})
+\frac{\epsilon_3}{\hbar}(\rho_{31}-\rho_{13})+\frac{J}{\hbar}
(\rho_{23}-\rho_{32}) \label{eqB} \ee
and
\be 0=\frac{J}{\hbar}(\rho_{13}-\rho_{31})+\frac{\epsilon_2-
\epsilon_3}{\hbar}(\rho_{23}-\rho_{32})-\frac{i}{2}\Gamma_{\downarrow}
(\rho_{23}+\rho_{32}). \label{eqC} \ee In combination with
(\ref{Ass3}), Eqs. (\ref{eqA}-\ref{eqC}) give
\be \rho_{12}=\rho_{21}=\frac{K/\hbar}{2\Gamma+\frac{1}{2}
\Gamma_{\uparrow}}\Im{\rm m}\ \rho_{31}, \label{eqD} \ee
\be \Re{\rm e}\ \rho_{31}=\frac{\epsilon_3}{2\hbar\Gamma}\Im{\rm m}
\ \rho_{31}, \label{eqE} \ee
and
\be \rho_{23}=\rho_{32}=-\frac{2J}{\hbar\Gamma_{\downarrow}}\Im{\rm m}
\ \rho_{31}. \label{eqF} \ee

Let us now take difference of the fourth and sixth equation in (\ref{Liou})
in the stationary state. Owing to (\ref{eqD}-\ref{eqE}), it gives
\be 0=-\frac{J}{\hbar}(\rho_{11}-\rho_{22})+\left\{-\frac{\epsilon_2}{\hbar}
\frac{2K/\hbar}{2\Gamma+\frac{1}{2}\Gamma_{\uparrow}}-\frac{K}{\hbar}
\frac{\epsilon_3}{\hbar\Gamma}\right\}\Im{\rm m}\ \rho_{31}. \label{eqI} \ee
Similarly, from the difference of the sixth and seventh equation and taking
into account (\ref{eqA}), (\ref{eqE}) and (\ref{eqF}), we obtain
\be 0=\left[(\frac{K}{\hbar})^2\frac{1}{\Gamma+\frac{1}{4}\Gamma_{\uparrow}}
+(\frac{J}{\hbar})^2\frac{4}{\Gamma_{\downarrow}}+(\frac{\epsilon_3}{\hbar}
)^2\frac{1}{\Gamma}+2(2\Gamma+\Gamma_{\uparrow}+\Gamma_{\downarrow})
\right]\Im{\rm m}\ \rho_{31}. \ee
As the expression in the square brackets is always positive, this implies
that
\be \Im{\rm m}\ \rho_{31}=0, \label{eqJ1} \ee
i.e. using (\ref{eqA}-\ref{eqC})
\be \rho_{13}=\rho_{31}=\rho_{12}=\rho_{21}=\rho_{23}=\rho_{32}=0.
\label{eqJ2} \ee
On the other hand, from (\ref{eqI}) and (\ref{eqJ1}), we get that in
the stationary state
\be \rho_{11}=\rho_{22}. \label{eqPom} \ee
One should realize that conditions (\ref{eqPom}) and $\rho_{12}=\rho_{21}$
(see (\ref{eqJ2})) obtained so far fully correspond to what has been said
above about no-flow equilibrium inside the dimer `1 - 2'.

The eighth or ninth equation of (\ref{Liou}) yield, in the
stationary state and with the help of (\ref{eqJ2}),
\be \rho_{22}-\rho_{33}=0. \label{eqK} \ee Together with
(\ref{norm}) and (\ref{Ass2}), it provides an inhomogeneous set of
three linear algebraic equations determining the site occupation
probabilities (all the time provided that the no-flow condition
(\ref{Ass1}) used above applies). The solution reads
\be \rho_{11}=\frac{\Gamma_{\downarrow}}{\Gamma_{\downarrow}+
2\Gamma_{\uparrow}}, \quad
\rho_{22}=\rho_{33}=\frac{\Gamma_{\uparrow}
}{\Gamma_{\downarrow}+2\Gamma_{\uparrow}}. \label{presol} \ee This
result, on the other hand, contradicts (\ref{eqPom}). This is the
required contradiction implying that (\ref{Ass1}) cannot be
correct. One can also ask what is the reason for the
contradiction. Clearly, (\ref{eqPom}) would be satisfied by
(\ref{presol}) if there were
$\Gamma_{\downarrow}-\Gamma_{\uparrow}=0$. That would, however,
mean to disregard the spontaneous processes that are responsible
for the difference on the left hand side. The spontaneous
processes are, however, purely quantum. Similarly, one can easily
observe that (\ref{presol}) becomes fully compatible with
(\ref{eqPom}) in the limit of the infinite temperature
$T_{II}\rightarrow+\infty$. The infinite temperature limit means,
however, the classical limit (the Bohr correspondence principle).
All that is why we can understand the violation of the second law
we arrive at below (as well as in other models yielding such a
striking conclusion - see above) as a consequence of quantum
effects.

So, there is always an electron circular flow in the system
implying (not in general but) in our specific situation nonzero
heat transfer ${\cal Q}$ (as given by (\ref{Qflo}) and
(\ref{flo1}) or (\ref{flo2})) between baths I and II. The last
questions to be solved before we resort to a numerical study are
what is its orientation and how the conclusion contradicts the
second law.

\section{Violation of the second law}

In order to infer what is the orientation of the mean heat flow,
let us turn to above formulae (\ref{flo1}) and (\ref{Qflo}). From
(\ref{Qflo}) we get that signs of ${\cal Q}$ and ${\cal J}$
coincide ($\epsilon_3>0$). As for the latter, we remind that
$\rho_{33}$ is always (as a site occupation probability) positive
and that $\Gamma_{\uparrow}$ disappears for $T_{II}\rightarrow0$.
Thus, from (\ref{flo1}), we get that ${\cal J}$ is, in the low
temperature limit of bath II but arbitrary nonzero $T_I$, always
positive. This is, by the way, also what our numerical results
show.

Let us now increase $T_{II}$. One should realize that ${\cal Q}$
is a continuous function of $T_{II}$ and never turns to zero. (For
that, see the above proof.) So it should remain positive even when
$T_{II}$ becomes greater than $T_I$. (In fact, owing to
intermixture of $T_I$ with details of coupling to bath I in
$H_{S-B}^I$ inside $\Gamma$, nothing can happen at the moment when
$T_{II}$ passes $T_I$. This fact was also confirmed numerically.)
Positive values of ${\cal Q}$ mean, however, a positive rate of
heat transfer from bath I to bath II which thus goes, for
$T_I<T_{II}$, {\em against} temperature step. As the heat transfer
is spontaneous (there is no external expenditure of energy or
whatever else conditioning this transfer), this conclusion
explicitly contradicts the Clausius form of the second law
\cite{Clau} stating that such processes are impossible. On the
other hand, the conclusion obtained analytically here (and
verified numerically below) that the second law is in our system
really violated fully corresponds to conclusions of \cite{Shee1}
where, for an experimental plasma system corresponding to the
above model, the universal validity of the second law was first
seriously challenged.

\section{Numerical results}

In order to verify the above conclusions, we have solved the set
(\ref{Liou}) and (\ref{norm}) numerically. There was also a
secondary reason for this numerical study: Analytically, we were
unable to prove that the heat transfer really turns to zero in the
limit of zero temperature $T_I$ of bath I. This is what must be
expected physically because in such a limit, there is no heat
available  in bath I to be transferred to bath II. In just other
words: There is no dephasing in this limit between sites 1 and 2.
So, the covalent bond 1-2 should become perfect, making thus the
electron (and consequently also the heat) flow impossible.

Fig. 1 shows typical results. We designate
$\gamma_0=\Gamma_{\uparrow}[{\rm e}^{\beta_{II}\epsilon_3}-1]$.
Three things are worth noticing:
\begin{itemize} \item In accordance with the above analytical
arguments, the mean heat flow ${\cal Q}$ is always positive (i.e.
going from bath I to bath II).
\item With decreasing dephasing rate $2\Gamma$ corresponding to
decreasing temperature $T_I$, ${\cal Q}$ turns apparently to
zero.
\item For constant rate $2\Gamma$, ${\cal Q}$ is only very little
dependent on temperature $T_{II}$ of bath II. Slight increase as
well as decrease with $T_{II}$ are both possible. This may be
interpreted as a result of two competing tendencies:
\begin{itemize} \item Increasing $T_{II}$ increases also the rate
of dephasing between sites 2 and 3 caused by nonzero and
$T_{II}$-dependent terms $-\frac{i\hbar}{2}\Gamma_{\downarrow}$ in
3-3 and 4-4 elements of block ${\cal C}$ in (\ref{Liou}).
Similarly the terms $-\frac{i\hbar}{2}\Gamma_{\uparrow}$ in 4-4
and 5-5 elements of block ${\cal A}$ in (\ref{Liou}) contributing
to dephasing of sites 1 and 2. This leads to greater violations of
the 2-3  and 1-2 bonds, i.e. to increase of ${\cal Q}$. \item
Increasing $T_{II}$ on the other hand implies relatively
decreasing role of the spontaneous processes in $3\rightarrow1$
transitions what means suppression of ${\cal Q}$.
\end{itemize} Dependence of ${\cal Q}$ on $T_{II}$ is, however,
always very small.
\end{itemize}

\section{Conclusion}

We have obtained a spontaneous heat flow between two {\em
macroscopic} baths that is owing to a specific activity of our
microscopic single-electron system not aided from outside. So,
starting from rigorous mathematics of the quantum theory of open
systems, a contradiction with the second law of thermodynamics has
been obtained for the model in question. Remind that except for
model assumptions, no approximations were made that could be made
responsible for the effect, and that the model corresponds to an
experimental system positively tested in \cite{Shee1}. This
indicates that one should choose between just two alternatives:
\begin{itemize} \item There is still something hidden in physics
beyond the second law what is at present not fully understood.
This possibility might also mean complementing contemporary
quantum mechanics and present philosophy of quantum-mechanical
modelling in order to reconcile the quantum theory with
(presumably) universally valid thermodynamics.
\item The second alternative is to refrain from the so far universally
assumed validity of thermodynamics in the macroworld. One should
realize that though our system is microscopic, appending the
macroscopic reservoirs turn the physics to the macroscopic one.
\end{itemize}

\section{Acknowledgement}
The authors should like to acknowledge support from grants
202/99/0182 of the Czech grant agency and 153/99/B of the Grant
agency of Charles University, Prague. V. \v{C}. should also like
to acknowledge discussions with D. P. Sheehan that he had during
his visit to San Diego. Their previous collaboration lead to paper
\cite{CapShe} where similar ideas (though still without
mathematical details) first appeared.

\section{Appendix: Derivation of the starting equations by non-scaling arguments}

Arguments in favour of violation of the second law as reported
above critically depend on the existence of the heat flow from
bath I to bath II, i.e. on the existence of the electron mean
circular flow ${\cal J}$. Existence of the flow has been above
proved using scaling which does not belong to a generally accepted
weaponry in kinetic theories. Moreover, though standard practice
confirms such a possibility, one could ask about justification to
use such a (as well as any other) kinetic approach beyond (time)
limits of the kinetic regime. (Notice, e.g., that constant $a$ in
$...\sup_{0\le\lambda^2t\le a}...$ in (\ref{Davres}) is always
finite.) That is why we present another treatment below that is
non-scaling but fully confirms the above conclusions. This
treatment is then fully resistive even against such objections as
it assumes, as experiments require, taking first (though after the
thermodynamic limit of the bath) the dc, i.e. the infinite time
limit. {\em Only then} (if at all) discussion based on smallness
of individual terms in the Hamiltonian (coupling constants etc.)
comes into question.

For reasons connected with unreliability of the finite-order
approximations in convolution theories discussed in \cite{CapPhA},
we refrain from convolution theories of the Nakajima-Zwanzig
\cite{Naka,Zwan} type. Instead, we adopt the formalism based on
the time-convolutionless (i.e. time-local) Generalized master
equations (TCL-GME). These were first suggested by Fuli\'{n}ski
and Kramarczyk \cite{Fuli,FulKra} but more operative (and in fact
equivalent \cite{Gzyl}) are those by Shibata, Hashitsume,
Takahashi, and Shingu \cite{HaShSh,ShTaHa}. The starting Shibata,
Hashitsume, Takahashi, and Shingu identity (derived as a direct
consequence of the Liouville equation) reads
\[ \frac{d}{dt}{\cal P}\rho^{S+B}(t)=-i{\cal PL}[1+i\int_0^t
\exp\{-i(1-{\cal P}){\cal L}\tau\}(1-{\cal P}){\cal
LP}\exp\{i{\cal L}\tau\}\,d\tau]^{-1}\] \be\cdot [\exp\{-i(1-{\cal
P}){\cal L}t\}(1-{\cal P})\rho^{S+B}(0)+{\cal P}\rho^{S+B}(t)].
\label{SHTS} \ee Now we make two steps. First we take for ${\cal
P}$ the Argyres-Kelley projector (\ref{ArgKel}), and assume the
initial condition (\ref{inicond}). Let us now split our
Hamiltonian as in (\ref{split}) and introduce ${\cal L}_0$ and
$\lambda{\cal L}_1$ as in (\ref{Livspl}). This reduces, for both
the weak coupling (\ref{WCchoi}) and our identification
(\ref{CapBarid}), equation (\ref{SHTS}) to
\be \frac{d}{dt}\rho(t)=(-i{\cal L}_0-i\langle{\cal
L}_1\rangle+\lambda^2{\cal K}^{TCL-GME}(t))\rho(t). \label{Lioeq2}
\ee Here, because ${\cal PL}_0={\cal L}_0{\cal P}$,
\[ \lambda^2{\cal K}^{TCL-GME}(t)\ldots=-Tr_B\left({\cal L}_1[1+i\int_0^t
\exp\{-i(1-{\cal P}){\cal L}\tau\}(1-{\cal P}){\cal L}_1{\cal
P}\exp\{i{\cal L}\tau\}\,d\tau]^{-1}\right.\]
\be\left.\cdot\int_0^t \exp\{-i(1-{\cal P}){\cal L}\tau\}(1-{\cal
P}){\cal L}_1{\cal P}\exp\{i{\cal
L}\tau\}\,d\tau(\rho^B\otimes\ldots)\right). \label{NewK} \ee
Notice that the relaxation superoperator $\lambda^2{\cal
K}^{TCL-GME}(t)$ still involves also higher-than-second order
terms in $\lambda$. It is also time-dependent. This time
dependence is determined by the decay-to-zero of the integrand
(after the implicit but everywhere assumed thermodynamic limit of
the bath). The characteristic times are given by dephasing in the
bath. Once we disregard a transient (and for our purposes fully
unimportant) initial time period, we may turn time $t$ in
(\ref{NewK}) to infinity.

We should like to stress that \begin{itemize} \item turning $t$ to
infinity in $\lambda^2{\cal K}^{TCL-GME}(t)$ still does not mean
treating just the final result of the finished process of
relaxation;
\item the long-time limit taken in (\ref{NewK}) is dictated by the
experimental situation (we are not interested in short-time
transient relaxation effects) and time $t$ could be, beyond some
bath dephasing time, well taken arbitrarily large, in accordance
with the experiment. \item This is unlike the situation with the
value of $\lambda$. In Nature, this coupling constant is
determined, for a given experiment, once for ever and cannot be
arbitrarily changed. \end{itemize} That is why one should expand,
if at all, in powers of $\lambda$ only {\em after} the long time
limit $t\rightarrow+\infty$ in $\lambda^2{\cal K}^{TCL-GME}(t)$.
That is what we do here. If we take $t\rightarrow+\infty$ in
(\ref{NewK}) and then limit our attention to the lowest
nonvanishing (i.e. the second) order in $\lambda$, (\ref{NewK})
then turns to
\[ \lambda^2{\cal K}^{TCL-GME}(t)\ldots\] \be=-Tr_B\left({\cal
L}_1\int_0^{+\infty} \exp\{-i{\cal L}_0\tau\}(1-{\cal P}){\cal
L}_1{\cal P}\exp\{i{\cal L}_0\tau\}\,d\tau(\rho^B\otimes\ldots)
\right)+{\cal O}(\lambda^4). \label{NewK2} \ee The first term on
the right hand side of (\ref{NewK2}) should now be compared with
that of $\lambda^2{\cal K}$ in (\ref{Davres}). The point is that
if identity of these two expressions for the relaxation
superoperator is established, the TCL-GME (\ref{Lioeq2}) becomes,
for both the standard weak coupling choice (\ref{WCchoi}) and that
one by \v{C}\'{a}pek and Barv\'{\i}k (\ref{CapBarid}), fully
equivalent to (\ref{Lioeq}).

At the first sight, there is a similarity but no identity
observed. In order to discuss the point in detail, let us make
several physically motivated and justifiable steps. First, the
integrations, in both $\lambda^2{\cal K}$ in (\ref{Davres}) and
$\lambda^2{\cal K}^{TCL-GME}(t)$ in (\ref{NewK}), lead (upon
explicit introduction of the matrix elements involved) to
distributions $\lim_{\delta\rightarrow0+}\frac{-i\hbar}{\Delta
E-i\hbar\delta}$ that we approximate as $\pi\hbar\delta(\Delta E)$
($\Delta E$ being relevant differences of eigenenergies of $H_0$).
This step amounts to neglect term v.p.$\frac{-i\hbar}{\Delta E}$
(here v.p. means the fraction in the Cauchy sense); such imaginary
terms are standardly interpreted as just renormalizations of
transfer (hopping or resonance) integrals. (In this connection,
notice the imaginary unit $i$ in the $(-i{\cal L}_0-i\langle{\cal
L}_1\rangle)$ term on the right hand side of (\ref{Lioeq2}) or
(\ref{Davres}).) Doing so, the $mn$-matrix element of
$\lambda^2{\cal K}\ldots$ in (\ref{Davres}) turns to
\[\left(\lambda^2{\cal K}\ldots\right)_{mn}=-\frac{1}{\hbar}
\sum_{\mu}\sum_{s\sigma}\sum_{q\kappa}\{\langle m\mu|\lambda
H_1|s\sigma\rangle\langle s\sigma|\lambda H_1|q\kappa\rangle
\langle q\kappa|\rho^B\otimes\ldots|n\mu\rangle\pi
\delta(E_{s\sigma}-E_{m\mu})\]
\[-\langle m\mu|\lambda H_1|s\sigma\rangle\langle s\sigma|\rho^B
\otimes\ldots|q\kappa\rangle\langle q\kappa|\lambda
H_1|n\mu\rangle\pi\delta(E_{s\sigma}-E_{m\mu})\]
\[ -\langle m\mu|\lambda H_1|q\kappa\rangle\langle
q\kappa|\rho^B\otimes\ldots|s\sigma\rangle\langle s\sigma|\lambda
H_1|n\mu\rangle\pi\delta(E_{n\mu}-E_{s\sigma})\] \[+\langle
m\mu|\rho^B\otimes\ldots|q\kappa\rangle\langle q\kappa|\lambda
H_1|s\sigma\rangle\langle s\sigma|\lambda
H_1|n\mu\rangle\pi\delta(E_{n\mu}-E_{s\sigma})\}\] \be
+\left(\lambda^2\Delta{\cal K}\ldots\right)_{mn}.
\label{Davrelsup} \ee Here, we have used the notation
$|m\mu\rangle=|m\rangle\otimes|\mu\rangle$ etc. where $|m\rangle$
and $|\mu\rangle$ are respectively eigenstates of $H_0-H_B$ and
$H_B$; $E_{m\mu}=E_m+E_{\mu}$ are the corresponding eigenenergies.
In the same way, disregarding already the transient
time-dependence and higher-than-second order terms ({\em by the
definition} absent in (\ref{Davres}))
\[\left(\lambda^2{\cal K}^{TCL-GME}\ldots\right)_{mn}=-\frac{1}{\hbar}
\sum_{\mu}\sum_{s\sigma}\sum_{q\kappa}\{\langle m\mu|\lambda
H_1|s\sigma\rangle\langle s\sigma|\lambda H_1|q\kappa\rangle
\langle q\kappa|\rho^B\otimes\ldots|n\mu\rangle\pi
\delta(E_{s\sigma}-E_{q\kappa})\]
\[-\langle m\mu|\lambda H_1|s\sigma\rangle\langle s\sigma|\rho^B
\otimes\ldots|q\kappa\rangle\langle q\kappa|\lambda
H_1|n\mu\rangle\pi\delta(E_{q\kappa}-E_{n\mu})\]
\[ -\langle m\mu|\lambda H_1|q\kappa\rangle\langle
q\kappa|\rho^B\otimes\ldots|s\sigma\rangle\langle s\sigma|\lambda
H_1|n\mu\rangle\pi\delta(E_{m\mu}-E_{q\kappa})\] \[ +\langle
m\mu|\rho^B\otimes\ldots|q\kappa\rangle\langle q\kappa|\lambda
H_1|s\sigma\rangle\langle s\sigma|\lambda
H_1|n\mu\rangle\pi\delta(E_{q\kappa}-E_{s\sigma})\}\] \be
+\left(\lambda^2\Delta{\cal K}^{TCL-GME}\ldots\right)_{mn}.
\label{TCLrelsup} \ee Obviously, the second and third terms from
(\ref{Davrelsup}) equal respectively to the third and second terms
in (\ref{TCLrelsup}). As for the remaining terms, there are
differences between the Davies (\ref{Davrelsup}) and TCL-GME
(\ref{TCLrelsup}) results that may become important when, e.g.,
different matrix elements of the coupling to phonons interfere.

In order to show that this is not our case for the above model and
the regime investigated, let us make several specifications and
observations. \begin{itemize} \item First, assume that
\be \langle\mu|\rho^B|\nu\rangle=\delta_{\mu,\nu}p_{\mu}.
\label{diagcon} \ee This is consistent with assumptions od both
the scrutinized approaches.
\item Realize that for the model specified by
(\ref{H_S}-\ref{H_S-B}), the two terms in the system-bath coupling
then do not interfere. For the site-local coupling, it is trivial
to see the equivalence of the first and the fourth terms in
(\ref{Davrelsup}) and (\ref{TCLrelsup}). So, we turn our attention
to just the site off-local coupling term $H_{S-B}^{II}$ in
(\ref{H_S-B}). \item We specify our reasoning here to the regime
corresponding to our choice (\ref{CapBarid}). Then $H_S|_{J=K=0}$
is site-diagonal, i.e. the Latin summation indices in
(\ref{Davrelsup}) and (\ref{TCLrelsup}) are sites.
\end{itemize} Because of the last point, equivalence between the
first rows of (\ref{Davrelsup}) and (\ref{TCLrelsup}) and,
similarly, between the fourth rows of  (\ref{Davrelsup}) and
(\ref{TCLrelsup}), is then for the site off-local coupling easily
seen. [For example, with $\lambda H_1$ being in the first row of
(\ref{Davrelsup}) and (\ref{TCLrelsup}) substituted by
$H_{S-B}^{II}$ from (\ref{H_S-B}), the form of $H_{S-B}^{II}$
implies that nonzero contribution appears just for $m=q$. Because
of (\ref{diagcon}), only terms with $\kappa=\mu$ contribute what
makes the equivalence between the first rows of (\ref{Davrelsup})
and (\ref{TCLrelsup}) explicit.] As for the last terms in
(\ref{Davrelsup}) and (\ref{TCLrelsup}) ignored so far, we get
\[ \left(\lambda^2\Delta{\cal
K}\ldots\right)_{mn}=\int_0^{+\infty}dx\,Tr_B\ ({\rm e}^{i{\cal
L}_0x}{\cal L}_1{\rm e}^{-i{\cal L}_0x}{\cal P}{\cal
L}_1(\rho^B\otimes...))\]
\[=\frac{1}{\hbar}\sum_{\mu\nu\alpha}\sum_a\{\langle
m\mu|\lambda H_1|a\alpha\rangle\langle\alpha|\rho^B|\mu\rangle
\langle a\nu|[\lambda
H_1,\rho^B\otimes\ldots]|n\nu\rangle\pi\delta(
E_{a\alpha}-E_{m\mu})\] \be-\langle m\nu|[\lambda
H_1,\rho^B\otimes\ldots]|a\nu\rangle\langle\mu|
\rho^B|\alpha\rangle\langle a\alpha|\lambda
H_1|n\mu\rangle\pi\delta(E_{n\mu}-E_{a\alpha})\}. \label{corDav}
\ee and
\[ \left(\lambda^2\Delta{\cal K}^{TCL-GME}\ldots\right)_{mn}=
\int_0^{+\infty}d\tau\,Tr_B\left({\cal L}_1{\rm e}^{-i{\cal
L}_0\tau}{\cal P L}_1{\cal P}{\rm e}^{i{\cal
L}_0\tau}(\rho^B\otimes\ldots)\right)+{\cal O}(\lambda^4)
\] \[=\frac{1}{\hbar}\int_0^{+\infty}dx\sum_{\mu\nu\alpha\beta}\sum_{ab}[ \langle
m\mu|\lambda H_1|a\alpha\rangle\langle\alpha|{\rm
e}^{-iH_Bx/\hbar}\rho^B{\rm e}^{iH_Bx/\hbar}|\mu\rangle\]
\[\times\{\frac{-i\langle a\nu|\lambda H_1|b\beta\rangle\langle
b\beta|\rho^B\otimes\ldots|n\nu\rangle}{E_{a\nu}-E_{b\beta}-i0+}+
\frac{i\langle a\nu|\rho^B\otimes\ldots|b\beta\rangle\langle
b\beta|\lambda H_1|n\nu\rangle}{E_{b\beta}-E_{n\nu}-i0+}\}\]
\[-\{\frac{-i\langle m\nu|\lambda H_1|b\beta\rangle\langle
b\beta|\rho^B\otimes\ldots|a\nu\rangle}{E_{m\nu}-E_{b\beta}-i0+}
+\frac{i\langle m\nu|\rho^B\otimes\ldots|b\beta\rangle\langle
b\beta|\lambda H_1|a\nu\rangle}{E_{b\beta}-E_{a\nu}-i0+}\}\] \be
\cdot\langle\mu|{\rm e}^{-iH_Bx/\hbar}\rho^B{\rm
e}^{iH_Bx/\hbar}|\alpha\rangle\langle a\alpha|\lambda
H_1|n\mu\rangle] +{\cal O}(\lambda^4). \label{corTC} \ee Clearly,
expressions (\ref{corDav}) and (\ref{corTC}) are discernibly
different even when we omit the renormalization terms.
Fortunately, with (\ref{diagcon}) and the fact that both the terms
in $H_{S-B}$ in (\ref{H_S-B}) are off-diagonal in the phonon
indices, both (\ref{corDav}) and (\ref{corTC}) are in fact exactly
zero. This makes the proof of full equivalency, for our model and
to the lowest perturbational order, of the relaxation
superoperator ${\cal K}$ as derived from the scaling Davies theory
with that one derived by the non-scaling time-convolutionless
Generalized Master Equation theory complete. The latter theory
thus provides independent non-scaling way to our above equation
(\ref{Liou}) with (\ref{matr}) forming basis of the above
discussion. The important point is that we have here, after taking
the thermodynamic limit of the bath, first turned the real
physical time behind dephasing time of the bath, i.e. potentially
even to infinity. Only then we have discussed the form of the
relaxation tensor determining the relaxation process up to
infinite times as valid for small couplings. Hence, unlike the
scaling theories, strength of the coupling never comes into any
competition with time limitations of the theory so that no
objections can be raised that the kinetic theory used is
inapplicable behind some long critical times.

\newpage


\newpage

\begin{center}
{\bf Figure captions}
\end{center}

\noindent Figure 1: Spontaneous energy flow ${\cal Q}$ in units
$4K^2/\hbar$ from bath I to bath II as a function of temperature
$T_{II}=1/(k_B\beta_{II})$ of bath II. We set $J=K=0.5$ eV,
$\epsilon_2/K=4$, $\epsilon_3/K=2$, $\hbar\gamma_0/K=0.02$, and
$\hbar\Gamma/K=10^{-2},\,10^{-3}$ and $10^{-4}$ (decreasing
dephasing rate, i.e. also heating, at site 1) for curves a), b),
and c), respectively. Notice that $2\Gamma$ incorporates also
temperature $T_I$ of bath I.
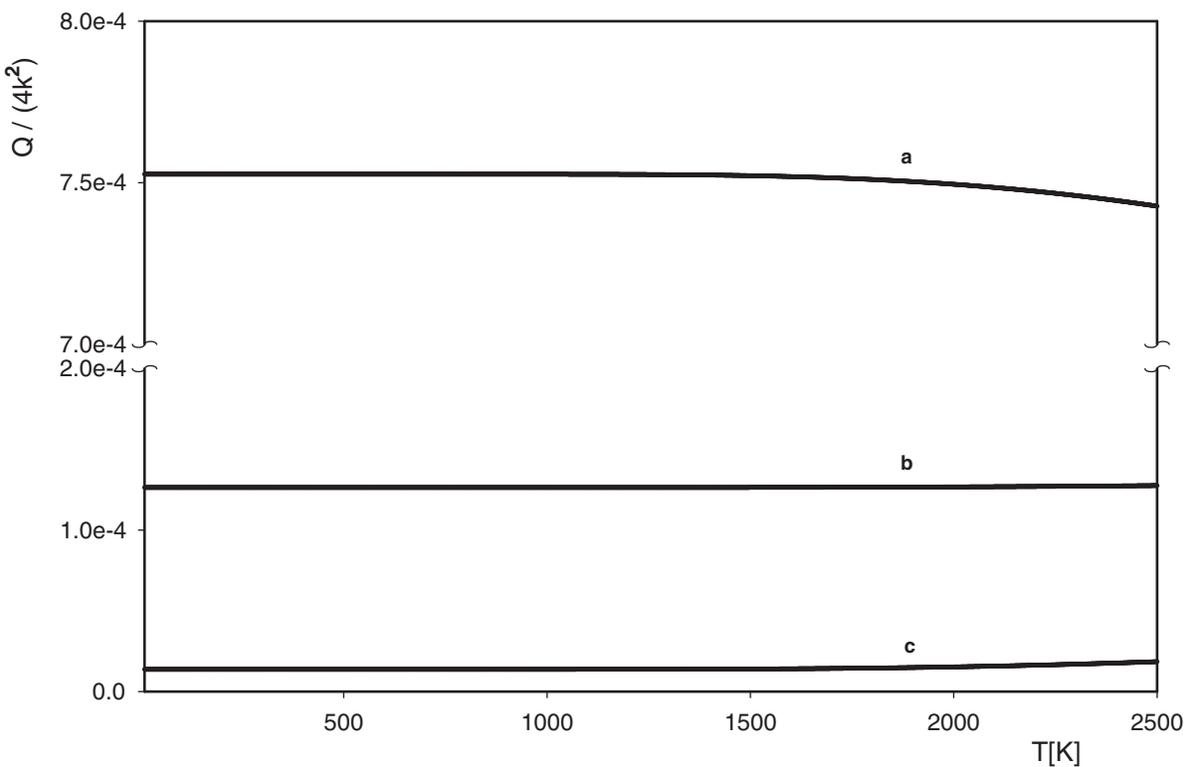
\end{document}